\documentclass{iopart}
\usepackage{iopams}
\usepackage{cite}
\usepackage{graphicx}
\usepackage[applemac]{inputenc}

\begin{document}
\title{Influence of higher-order harmonics on the saturation of the tearing mode}
\author{N.~Arcis, N.F.~Loureiro and F.~Militello}
\address{EURATOM/UKAEA Fusion Association, Culham Science Centre, Abingdon, Oxon, OX14 3DB, UK}
\ead{Nicolas.Arcis@ukaea.org.uk}

\begin{abstract}
The nonlinear saturation of the tearing mode is revisited in slab geometry by taking into account higher-order harmonics in the outer solution. The 
general formalism for tackling this problem in the case of a vanishing current gradient at the resonant surface is derived. It is shown that, 
although the higher-order harmonics lead to corrections in the final saturation equation, they are of higher order in the perturbation parameter, 
which provides a formal proof that the standard one-harmonic approach is asymptotically correct.
\end{abstract}


\section{Introduction}
The tearing mode \cite{Furth} is a resistive magnetohydrodynamic (MHD) instability that resonates on magnetic surfaces where the mode's 
wave vector is perpendicular to the magnetic field. It leads to the formation of so-called magnetic islands around the resonant surface, 
resulting in a local change of magnetic topology and a greater radial transport on a length scale of the order of the islands' full width, $w$. 
Therefore, these structures are of great interest for nuclear fusion, and it is especially important to have a keen insight into their 
stability properties as well as their maximum achievable amplitude.

While the linear theory of the tearing mode is rather well understood, the nonlinear one, pioneered by Rutherford 
thirty-five years ago \cite{Ruth}, is taking longer to unravel. Recently, a series of works have derived solutions for the saturation of the 
tearing mode in the framework of simple physical models \cite{MP,EO,Hastie,Arc05,Arc06,Militello}, somewhat rekindling interest in 
this subject. Although rigorous in their mathematical details, they share a common feature based on an assumption first made by 
Rutherford, i.e. they neglect higher-order (poloidal) harmonics in the magnetic perturbation. However, this is not really justified \emph{a 
priori}, since nonlinearities naturally couple all harmonics, and one may therefore question the validity of the afore-mentioned references.

In this work, we investigate this problem using the simplest of physical models, namely that of reduced MHD in slab geometry in 
the so-called symmetric case (i.e. no current gradient at the resonant surface), which is the same as that used in \cite{MP,EO}. We first 
derive the general solution to this problem and then explicitly solve the case with two harmonics for two different types of equilibrium. Last, 
we compare our results with those obtained from previous theories and draw conclusions.

\section{Model equations}

We first introduce the following normalizations:
\begin{equation}\label{eq:norm}
t=\tau_{\eta}\widetilde{t}\mbox{ ; }x=L\widetilde{x}\mbox{ ; }y=L\widetilde{y}\mbox{ ; }J=J_{N}\widetilde{J}\mbox{ ; }
\psi=\mu_{0}J_{0}L^2\widetilde{\psi}\mbox{ ; }\varphi=\frac{\eta}{\mu_{0}}\widetilde{\varphi},
\end{equation}
where $t$ is the time variable, $\tau_{\eta}=\mu_{0}L^2/\eta$ is the resistive diffusion time, $\mu_{0}$ is the permeability of free 
space, $\eta$ is the (uniform) resistivity, $x$ and $y$ are the radial and poloidal variables respectively, $L$ is a characteristic radial length, 
$J_{N}=J_{eq}(0)$, $J$ (resp. $J_{eq}$) is the (resp. equilibrium) current density, $\psi$ is the magnetic flux function (i.e. $\mathbf{B}
\equiv B_{z}\mathbf{e_{z}}+\nabla\times (\psi\,\mathbf{e_{z}})$, where $\mathbf{B}$ is the magnetic field and 
$\mathbf{e_{z}}$ is a unit vector perpendicular to the $xy$ plane) and $\varphi$ is the electric potential and plays the role of the (ion) 
stream function (i.e. $\mathbf{v}\equiv\mathbf{e_{z}}\times\nabla\varphi$, where $\mathbf{v}$ is the velocity field). Note that 
these normalizations are such that it is the equilibrium current density, and \emph{not} the equilibrium magnetic field, that is normalized to 
unity at the resonant surface. We then use the reduced MHD equations in slab geometry \cite{Strauss} which, taking into account the 
normalizations while omitting the ‘‘ $\widetilde{\;}$ '' for clarity, read:
\begin{equation}\label{eq:mouvement}
\partial_{t}\Delta_{\bot}\varphi+[\varphi,\Delta_{\bot}\varphi]=S^2[J,\psi]+
\frac{S}{Re}\Delta_{\bot}^2\varphi
\end{equation}
\begin{equation}\label{eq:Ohm}
\partial_{t}\psi+[\varphi,\psi]=J_{eq}-J
\end{equation}
\begin{equation}\label{eq:Ampere}
J=-\Delta_{\bot}\psi,
\end{equation}
where $S=v_{A}L\mu_{0}/\eta$ is the Lundquist number, $Re=v_{A}L/\nu$ is the Reynolds number, 
$v_{A}=J_NL\sqrt{\mu_{0}/\rho}$ is the Alfvén speed, $\nu$ is the viscosity and $\rho$ is the mass density ($\nu$ and $\rho$ are 
assumed to be constant). The Poisson brackets are given by $[f,g]\equiv \partial_{x}f\partial_{y}g-\partial_{x}g\partial_{y}f$. 
Finally, the resonant surface is conveniently set at the origin by letting $\psi_{eq}'(0)=0$, where the equilibrium magnetic flux function 
satisfies $\psi_{eq}''=-J_{eq}$.

\section{Perturbed equilibrium}

Provided there are no equilibrium flows in the $xy$ plane and given the fact that $S,Re\gg 1$, as is the case in present-day 
tokamak plasmas, the equation of motion (\ref{eq:mouvement}) simply yields:
\begin{equation}\label{eq:temp}
[J,\psi]=0.
\end{equation}
We then look for a perturbed equibrium of the form:
\begin{equation}\label{eq:perturbation}
\psi\sim\psi_{eq}(x)+\sum_{n\geq 1}\delta_{n}^2(t)\psi_{n}(x)\cos{\left[n\chi+\beta_{n}(t)\right]},
\end{equation}
where we have defined $\chi\equiv ky$, $k$ being the mode's wave-number and, throughout this paper, ‘$\sim$' has the meaning ‘equals 
plus higher order terms'. Note that, without loss of generality, it is possible to choose $\beta_{1}=0$, which will henceforth be the case. 
Substituting this expression into (\ref{eq:temp}), the $\psi_{n}$ satisfy:
\begin{equation}\label{eq:linear}
\psi_{n}''+\left[\frac{J_{eq}'}{\psi_{eq}'}-(nk)^2\right]\psi_{n}=0.
\end{equation}

At this point, we should clarify the case of the $n=0$ component. Indeed, to be fully general, one may be tempted to include a term 
of the form $\delta_{0}^2(t)\psi_{0}(x)$ in (\ref{eq:perturbation}) but, because of its Poisson bracket nature, (\ref{eq:temp}) is 
automatically satisfied in order $\delta_{0}^2$ and $\psi_{0}(x)$ remains unconstrained. Therefore, the $n=0$ component can only 
originate in Ohm's law's quasilinear terms, and one can show that this leads to an order $\sum_{n\geq 1}\delta_{n}^4$ change to the 
equilibrium, which, as was already noted in \cite{Ruth}, can be neglected in (\ref{eq:perturbation}). It has also been shown that the 
magnetic island itself may result in an $n=0$ component, but the latter has no impact whatsoever on the saturated island width 
\cite{Arc05,Arc06}. Since the main interest of the present paper is the saturation of the tearing mode, we shall ignore the $n=0$ 
component altogether for the sake of clarity.

In the following, we suppose that the first harmonic dominates the others at all times, i.e. that $\delta_{1}\geq \delta_{n>1}$. This assumption 
certainly makes sense in the linear regime, since the first harmonic is the most unstable one \cite{Furth}, but it also has to be consistent with the 
final saturation result, which will be shown to be the case later on. Based on the work already done in e.g. \cite{Arc05,Arc06}, one can then infer from 
(\ref{eq:Ohm}) that there is a boundary layer of width $\delta_{1}$ centered on $x=0$ and that one thus has to resort to the technique of 
asymptotic matching. Writing $J_{eq}\sim 1+a_{2}x^2$, choosing $\psi_{eq}(0)=0$ for convenience and solving (\ref{eq:linear}) using 
Fr\oe benius' method \cite{Jeff}, we derive the following expansion for the \emph{outer} solution:
\begin{eqnarray}\label{eq:outer}
\zeta_{out}&\sim&\frac{\xi^2}{2}-\sum_{n\geq 1}\alpha_{n}
\cos{(n\chi+\beta_{n})}\nonumber\\
& -&\delta_{1}|{\xi}|\sum_{n\geq 1}\alpha_{n}
\frac{\Delta_{n}'}{2}\cos{(n\chi+\beta_{n})}\\
& +&\delta_{1}^2\left\{a_{2}\frac{\xi^4}{12}-\sum_{n\geq 1}\alpha_{n}
\left[a_{2}+\frac{(nk)^2}{2}\right]\xi^2\cos{(n\chi+\beta_{n})}\right\}\nonumber ,
\end{eqnarray}
where we have defined $\zeta_{out}\equiv -\psi/\delta_{1}^2$, $\alpha_{n}\equiv(\delta_{n}/\delta_{1})^2$ and the inner 
variable $\xi\equiv x/\delta_{1}$ in order to make the ordering in the small parameter $\delta_{1}$ explicit. As to the linear stability 
parameters, $\Delta_{n}'$, they are related to the logarithmic jump of the $\psi_{n}$ around the resonant surface \cite{Furth}, namely
\begin{equation}
\Delta_{n}'=\lim_{\epsilon\rightarrow 0^{+}}[\psi_{n}'(\epsilon)-\psi_{n}'(-\epsilon)]/\psi_{n}(0),
\end{equation}
and depend on both the equilibrium current density profile and the boundary conditions.

The solution (\ref{eq:outer}) holds in the so-called outer region of the plasma where ideal MHD is a good approximation to our set 
of equations, but breaks down around the resonant surface where resistivity has to be taken into account. The solution that is valid in this 
resistive boundary layer is the \emph{inner} one which we derive in the next section. It utlimately has to be matched to (\ref{eq:outer}) that, in effect, is analogous to a 
boundary condition at infinity (i.e. $|\xi|\rightarrow\infty$). At this stage, it is perhaps worthwhile to stress once more that it is the $n>1$ 
terms in the sums appearing in (\ref{eq:outer}) that were neglected in previous works.

\section{Solution in the inner region}\label{sec:inner}
\subsection{Inner equations}

The inner equations are obtained by re-writing (\ref{eq:mouvement})-(\ref{eq:Ampere}) with respect to the inner variable $\xi$:
\begin{equation}\label{eq:GS}
[\zeta,J]=0
\end{equation}
\begin{equation}\label{eq:Ohmin}
-\delta_{1}^2\partial_{t}\zeta+\delta_{1}\partial_{t}{\delta}_{1}(\xi\partial_{\xi}\zeta-2\zeta)+k\delta_{1}[\zeta,\varphi]\sim
1+a_{2}\delta_{1}^2\xi^2-J
\end{equation}
\begin{equation}\label{eq:Amperein}
J=\partial_{\xi}^2\zeta+k^2\delta_{1}^2\partial_{\chi}^2\zeta,
\end{equation}
where the Poisson brackets are now taken with respect to the $(\xi,\chi)$ variables. Note that (\ref{eq:GS}) is valid in the nonlinear 
regime only, where the island is supposed to be larger than the resistive and visco-resistive layer widths \cite{Ruth}. This set of 
equations is then classically solved using perturbation expansions in powers of $\delta_{1}$, i.e. writing $\zeta=\sum_{l\geq 0}
\delta_{1}^{l}\zeta_{l}$, $\varphi=\sum_{l\geq 0}\delta_{1}^{l}\varphi_{l}$, and $J=\sum_{l\geq 0}\delta_{1}^{l}J_{l}$.

\subsection{Order $\delta_{1}^0$}
Ohm's law simply gives $J_{0}=1$, and the integration of (\ref{eq:Amperein}) together with the matching condition provided by 
(\ref{eq:outer}) at this order yields:
\begin{equation}\label{eq:zeta0}
\zeta_{0}=\xi^2/2-\sum_{n\geq 1}\alpha_{n}\cos{(n\chi+\beta_{n})}.
\end{equation}

\subsection{Order $\delta_{1}$}
Equation (\ref{eq:GS}) implies that the order $1$ component of $J$ is a function of $\zeta_{0}$ only on either side of the resonant 
surface, i.e. $J_{1}(\xi,\chi)=j_{1}(\zeta_{0};\pm)$, where we have defined $\pm\equiv {sign}(\xi)$. 
It is therefore easier to work in $(\zeta_{0},\chi;\pm)$ variables, which will be the case in the following. Ohm's law then yields:
\begin{equation}\label{eq:order1}
2\sum_{n\geq 1}\partial_{t}{\delta}_{n}\sqrt{\alpha_{n}}\cos{(n\chi+\beta_{n})}+k\xi\left.\partial_{\chi}\varphi_{0}
\right|_{\zeta_{0}}=-j_{1}(\zeta_{0};\pm).
\end{equation}
In order to solve this equation, it is convenient to define, for any function $f$, its flux average $\langle f\rangle$ as:
\begin{equation}
\left\{\begin{array}{l}
\displaystyle{\int_{-\pi}^{\pi} d\chi\ f(z,\chi;\pm)/\xi \quad\mbox{if } z\geq \zeta_{sep}\equiv
\zeta_{0}(0,\pi)}\\

\displaystyle{1/2\sum_{\sigma=\pm}\sigma\int_{-\chi_{0}}^{\chi_{0}}d\chi\ f(z,\chi;\sigma)/\xi \quad\mbox{if }  
z\leq \zeta_{sep}}
\end{array}\right. ,
\end{equation}
where $\chi_{0}\in [0,\pi]$ is the turning point of the corresponding flux surface, i.e. it satisfies $\zeta_{0}(0,\chi_{0})=z$, and, in this 
expression, $\xi$ has to be taken as a function of $(z, \chi;\pm)$, i.e. $\xi=\pm\sqrt{2}[z+\sum_{n\geq 1}\alpha_{n}\cos{(n\chi+\beta_{n})}]^{1/2}$. 
With this definition in mind, it is then easy to show that the solution to (\ref{eq:order1}) reads:
\begin{equation}\label{eq:j1}
j_{1}=-2\sum_{n\geq 1}\partial_{t}{\delta}_{n}\sqrt{\alpha_{n}}\left\langle\cos{(n\chi+\beta_{n})}\right\rangle/
\left\langle 1\right\rangle.
\end{equation}

\subsection{Order $\delta_{1}^2$}
The calculation is similar to that of the previous section except that we now neglect all terms dependant on $\partial_{t}{\delta}_{1}$ 
since they would only lead to higher order corrections in the final result. Consequently, (\ref{eq:Ohm}) gives:
\begin{equation}\label{eq:order2}
-\sum_{n\geq 1}\alpha_{n}\partial_{t}{\beta}_{n}\sin{(n\chi+\beta_{n})}+k\xi\left.\partial_{\chi}\varphi_{0}\right|_{\zeta_{0}}
=a_{2}\xi-j_{2}(\zeta_{0};\pm),
\end{equation}
which, after applying the bracket $\langle\cdots\rangle$ operator on both sides, yields:
\begin{equation}\label{eq:j2}
j_{2}=\left[ a_{2}\langle\xi^2\rangle+\sum_{n\geq 1}\alpha_{n}\partial_{t}{\beta}_{n}\langle\sin{(n\chi+\beta_{n})}\rangle
\right]/\left\langle 1\right\rangle.
\end{equation}

We stop the inner calculation here since it is the lowest relevant order. Indeed, we shall see in the next section that the asymptotic 
matching conditions on the $\Delta_{n}'$ terms in (\ref{eq:outer}) already provide a saturation theory for the tearing mode at that order.

\section{Asymptotic matching conditions}

When taking the asymptotic expansion of the inner solution derived in Section~\ref{sec:inner} for $|\xi|\rightarrow\infty$, one 
shows that the matching with the outer solution given by (\ref{eq:outer}) provides the following conditions (see, e.g., \cite{Arc06} for 
more details on the asymptotic matching procedure):
\begin{equation}\label{eq:matchingtemp1}
-\frac{2}{\pi}\int_{\zeta_{min}}^{+\infty}\frac{d\zeta_{0}}{\langle 1\rangle}\left\langle\cos{(m\chi+\beta_{m})}
\right\rangle E(\zeta_{0})\sim\alpha_{m}\Delta_{m}'
\end{equation}
\begin{equation}\label{eq:matchingtemp2}
\int_{\zeta_{min}}^{+\infty}\frac{d\zeta_{0}}{\langle 1\rangle}
\left\langle\sin{(m\chi+\beta_{m})}\right\rangle E(\zeta_{0})\sim0,
\end{equation}
where $\zeta_{min}\equiv\zeta_{0}(0,0)$ and $E(\zeta_{0})$ is given by:
\begin{equation}
E=j_{1}+\delta_{1}\sum_{n\geq 1}\alpha_{n}\left\langle 2a_{2}\cos{(n\chi+\beta_{n})}+\partial_{t}{\beta}_{n}
\sin{(n\chi+\beta_{n})}\right\rangle.
\end{equation}
Since we want to focus on the saturation of the mode, it is possible to simplify this set of 
equations. Indeed, (\ref{eq:matchingtemp2}) implies that, at saturation, $\beta_{n}\in\{0,\pi\}$ for all $n>1$ (recall that $\beta_{1}
=0$ by assumption). Therefore, we can do away with the $\beta_{n}$'s altogether provided we allow the $\alpha_{n}$'s to be either 
positive or negative (except for $\alpha_{1}$ that is always equal to $1$). If we do this, (\ref{eq:matchingtemp2}) is automatically met and, 
letting $\partial_{t}=0$, (\ref{eq:matchingtemp1}) can be re-written as:
\begin{equation}\label{eq:matching}
\alpha_{m}\Delta_{m}'+a_{2}\delta_{1}\sum_{n\geq 1}\alpha_{n}F_{mn}(\{\alpha_{i}\})\sim 0,
\end{equation}
where we have defined
\begin{equation}\label{eq:F}
F_{mn}(\{\alpha_{i}\})=\frac{4}{\pi}\int_{\zeta_{min}}^{+\infty}\frac{d\zeta_{0}}{\left\langle 1\right\rangle}
\left\langle\cos{m\chi}\right\rangle\left\langle\cos{n\chi}\right\rangle.
\end{equation}
Note that, trivially, $F_{mn}=F_{nm}$. The set of equations given by (\ref{eq:matching}) provides a general theory for the nonlinear 
saturation of the full island width $w_{s}\equiv 4\delta_{1}({\sum_{n\geq 0}\alpha_{2n+1}})^{1/2}$ when taking into account any 
number of harmonics in the outer solution (the island width is defined as the width of the separatrix at the O-point, where the 
separatrix is given by the equation $\zeta_{0}(\xi,\chi)=\zeta_{sep}\equiv\zeta_{0}(0,\pi)=\sum_{n\geq 1}(-1)^{n+1}\alpha_{n}$).

Although it is not possible to solve it analytically, we make an important comment about this result. Indeed, if we relax Rutherford's original 
assumption (namely $-\Delta_{n}'\gg 1$ for $n>1$) \cite{Ruth}, and instead (more reasonably) assume $\Delta_{n>1}'$ 
to be of order one, then (\ref{eq:matching}) implies $\alpha_{n>1}=O(a_{2}\delta_{1}/\Delta_{n}')=O(\delta_{1})$. 
Consequently, the higher-order harmonics only lead to a higher order correction in the saturation equation, i.e.:
\begin{equation}
\Delta_{1}'+a_{2}\delta_{1}\left(F_{11}+\sum_{n>1}\alpha_{n}F_{1n}\right)=\Delta_{1}'+a_{2}\delta_{1}F_{11}+
O(\delta_{1}^2)\sim 0,
\end{equation}
showing that the standard one-harmonic approach is asymptotically correct in the limit of vanishing $\delta_{1}$. Incidentally, the fact that 
$\alpha_{n>1}=O(\delta_{1})$ also ensures that our original assumption, namely that the first harmonic dominates over the higher ones, is 
consistent. In order to demonstrate 
this result quantitatively, we now illustrate our theory through the simplest case of two harmonics and compare the outcome with that 
of \cite{MP,EO}.

\section{Corrections due to the second harmonic for two specific equilibria}
Taking into account the first two harmonics only, (\ref{eq:matching}) gives:
\begin{eqnarray}
4\Delta_{1}'+a_{2}w_{s}(F_{11}+\alpha F_{12})&=&0\\
4\alpha\Delta_{2}'+a_{2}w_{s}(F_{21}+\alpha F_{22})&=&0,
\end{eqnarray}
where now, since we truncate after the second harmonic, $w_{s}=4\delta_{1}$ and we have written $\alpha=\alpha_{2}$ for the sake of clarity. Since $\alpha$ is expected to be small, it 
makes sense to Taylor expand the $F$ coefficients to first order as $F_{mn}\sim A_{mn}+\alpha B_{mn}$. It is then straightforward to 
solve for $w_{s}$ and $\alpha$:
\begin{equation}\label{eq:saturation}
w_{s}=\frac{-4\Delta_{1}'}{a_{2}\left[A_{11}+\alpha(B_{11}+A_{12})\right]}\mbox{ ; }
\alpha=\frac{A_{12}\Delta_{1}'}{\Delta_{2}'A_{11}-\Delta_{1}'(B_{12}+A_{22})},
\end{equation}
where $A_{11}\simeq 3.29$, $B_{11}\simeq 2.22$, $A_{12}\simeq 0.34$, $B_{12}\simeq 2.03$ and $A_{22}\simeq 1.59$ have been 
computed numerically.
\begin{figure}[bp]
\includegraphics[scale=0.535]{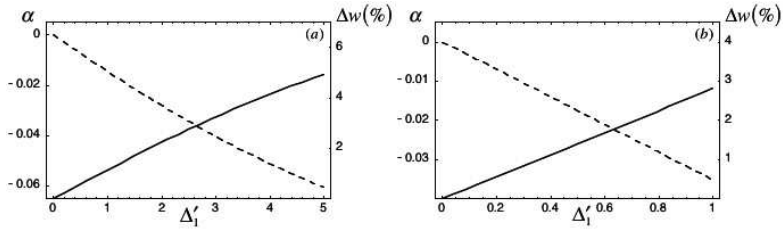}
\caption{\label{figure}Plot of $\Delta w$ (\full) and $\alpha$ (\dashed) as a function of $\Delta_{1}'$ for the cosh $(a)$ and sheet pinch $(b)$ 
equilibria. }
\end{figure}

These expressions can be evaluated provided $\Delta_{1}'$, $\Delta_{2}'$ and $a_{2}$ are given. To this end, we consider two 
specific equilibria often used in the literature: the cosh (i.e. $\psi_{eq}=1/(2\cosh^2{x})$ \cite{Porcelli}, where the $1/2$ factor has been 
included to comply with our normalization scheme) and sheet pinch (i.e. $\psi_{eq}=-\log{(\cosh{x})}$ 
\cite{Harris}) equilibria. The first one has $a_{2}=-4$ and the other $a_{2}=-1$, while the $\Delta_{n}'$ are given by:
\begin{equation}
\Delta_{n}'=\frac{2(5-n^2 k^2)(3+n^2 k^2)}{n^2 k^2\sqrt{4+n^2k^2}}\mbox{ and }
\Delta_{n}'=2\left(\frac{1}{nk}-nk\right)
\end{equation}
respectively. The results so-obtained are then compared with the one given by standard theory, namely $\hat{w}_{s}=-4\Delta_{1}'/
a_{2}A_{11}$ \cite{MP,EO}, and the outcome is shown in figure~\ref{figure}, where we have plotted $\Delta w\equiv (w_{s}-\hat{w}_{s})/
\hat{w}_{s}$ and $\alpha$ for both equilibria. We see that the island is found to be slightly bigger than was 
predicted by previous theory, but the corrections are very small, namely up to $5\%$ (resp. $3\%$) larger for the cosh (resp. sheet pinch) 
equilibrium with $\Delta_{1}'\lesssim   5$ (resp. $\Delta_{1}'\lesssim  1$). Of course, the greater $\Delta_{1}'$, the bigger these 
corrections can get, and, indeed, they reach up to $10\%$ (resp. $15\%$). However the constant-$\psi$ approximation breaks down for too 
large a $\Delta_{1}'$, so that the ranges shown in figure~\ref{figure} are, in effect, appropriate as far as our asymptotic matching theory is 
concerned. Therefore, we conclude that the standard theory of \cite{MP,EO} gives a correct result in its region of validity, which was 
expected since it has been confirmed numerically in \cite{Loureiro}.

\begin{figure}[bp]
\includegraphics[scale=1.2]{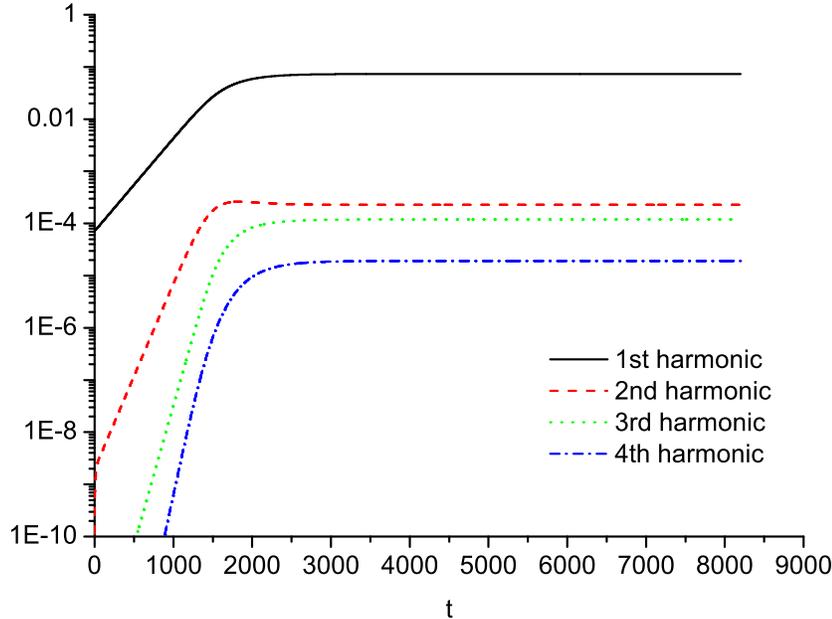}
\caption{\label{figure2}(colour online). Plot of the first four harmonics' amplitude versus time for the cosh equilibrium with 
$\Delta_{1}'$=1.0.}
\end{figure}

We now make one last remark concerning these results. One may naturally wonder whether the neglect of the third and higher harmonics is 
justified, since (\ref{eq:matching}) only implies that all $n>1$ harmonics are $O(\delta_{1}/\Delta_{n}')$. To test that this is the case, 
we have run a full numerical simulation for the cosh equilibrium with the pseudospectral code already used in \cite{Loureiro}, 
taking $\Delta_{1}'=1.0$. The results are shown in figure~\ref{figure2}, where we see that the second harmonic always has a greater amplitude 
than the third and fourth (as well as the higher-order ones, not shown). The reason for that is twofold: first, the nonlinear coupling to the first 
harmonic can be inferred to grow weaker for higher-order ones, and second, since $\Delta_{n}'\propto n$ for large $n$, the amplitude 
of the higher-order harmonics scales (at most) as $1/n$. Incidentally, we also see in figure~\ref{figure2} that the first harmonic largely 
dominates at all times, which gives further evidence that our original assumption (namely $\delta_{1}\geq\delta_{n>1}$) is correct, at 
least in the regime ($\Delta_{1}'$ not too large) we investigate.

\section{Conclusion}
In this paper, we have examined the impact of higher-order harmonics on the nonlinear evolution of the tearing mode in (symmetric) 
slab geometry. We have provided a general set of dynamical equations as well as a compact formula for the saturation of the mode. We have 
shown that the contribution due to the higher-order harmonics in the saturation equation led to a higher order correction in the perturbation 
parameter. Hence, we have justified the 
standard one-harmonic calculation as asymptotically valid. To make this claim more tangible, we have computed numerically the 
contribution due to the second harmonic for two types of equilibrium and shown that it indeed led to small corrections in a relevant range of 
values for $\Delta_{1}'$. Therefore, we believe this work usefully complements previous results in the theory of tearing mode saturation.

\ack
The authors would like to thank R.J.~Hastie for valuable comments. This work was funded by the United Kingdom 
Engineering and Physical Sciences Research Council, by the European Communities under the contract of Association between EURATOM 
and UKAEA, and by a EURATOM Intra-European Fellowship. The views and opinions expressed herein do not necessarily reflect those of 
the European Commission.

\end{document}